\documentclass[12pt,fleqn]{article}
\usepackage{psfrag,amsmath,epsfig,cite,times}
\newcommand{\F}{{\cal F}}

\newcommand{\VL}{\left( \begin{array}{c}}
\newcommand{\VR}{\end{array} \right)}

\def\pslash#1{{\setbox0=\hbox{$#1$}
  \rlap{\ifdim\wd0>.7em\kern.22\wd0\else\kern.1\wd0\fi /}#1}}

\newcommand{\dTB}{\delta t_\beta{}}

\allowdisplaybreaks[1]
\begin{document}
\begin{flushright}
DESY--02--181\\
{\tt hep-ph/0210372}\\
\end{flushright}
\vspace{3ex}
\begin{center}
{\Large\bf Gauge dependence and renormalization of $\tan\beta$\\}
\vspace{3ex}
{\large   Ayres Freitas,
          \underline{Dominik St{\"o}ckinger}
{\renewcommand{\thefootnote}{\fnsymbol{footnote}}
\footnote{
Talk presented by D.S. at the SUSY02 conference 
(10th International Conference on Supersymmetry 
and Unification of Fundamental Interactions), June 17--23 2002,
DESY, Hamburg, Germany. E-mail: {\tt dominik@mail.desy.de}.}}}
  \\[2ex]
  \parbox{10cm}{\small\center\em
           Deutsches Elektronen-Synchrotron DESY, 
  \\            D--22603 Hamburg, Germany
  }
\setcounter{footnote}{0}
\end{center}
\vspace{2ex}

\begin{abstract}
Renormalization schemes for $\tan\beta$ are discussed in view
of their gauge dependence. It is shown that several common
renormalization schemes lead to a gauge-dependent definition of
$\tan\beta$, whereas two classes of gauge-independent schemes show
even worse disadvantages. We conclude that the $\overline{DR}$-scheme
is the best compromise.
\end{abstract}

\section{Introduction}

The quantity $\tan\beta$ is one of the main input parameters of the
minimal supersymmetric standard model (MSSM). At tree-level it is
defined as the ratio of the two vacuum expectation values $v_{1,2}$ of
the MSSM Higgs doublets,
\begin{align}
\tan\beta & = \frac{v_2}{v_1}.
\label{TBDef}
\end{align}
Owing to its central appearance in the spontaneous symmetry breaking,
$\tan\beta$ plays a crucial role in almost all sectors of the MSSM and
has significant impact on most MSSM observables. 

As all parameters in
a quantum field theory, at higher orders $\tan\beta$ is actually
defined by the choice of a renormalization scheme. Though in principle
all renormalization schemes are equivalent, there are large practical
differences. The renormalization scheme determines the relation of
$\tan\beta$ to observables and thus its numerical value as well as its
formal properties like gauge dependence and renormalization-scale
dependence.

This talk is based on the analysis of \cite{TB}, where several
renormalization schemes for $\tan\beta$ were studied with the aim to
find an optimal scheme. One important aspect we will consider is the
gauge dependence induced by the renormalization schemes. Generally,
the dependence on the gauge fixing always drops out in relations
between different observables, but the relations between observable
quantities and formal input parameters like $\tan\beta$ can be
gauge dependent. As we will see, in some renormalization schemes,
$\tan\beta$ indeed turns out to be gauge dependent. 

Apart from gauge dependence there are two more desirable properties of
renormalization schemes: numerical stability of the perturbative
expansion, and process independence (in order not to spoil the
intuition that $\tan\beta$ is a universal quantity of the Higgs
sector). In the following we will discuss well-known and new schemes
in view of these properties, providing arguments that no ideal scheme
exists. The $\overline{DR}$-scheme will emerge as the best
compromise.

\section{Gauge dependence of well-known schemes}

Two classes of well-known and commonly used renormalization schemes
are the $\overline{DR}$-scheme and the schemes introduced in
\cite{Dabelstein,Rosiek}. At the one-loop level they are defined by
the conditions
\begin{align}
&\overline{DR}:& \dTB &\stackrel{!}{=} \mbox{pure
divergence},
\label{DRBarCond}
\\
&\mbox{DCPR}:& \dTB & \stackrel{!}{=}
\frac{1}{2 c_\beta^2 M_Z} {\rm Re}\Sigma_{A^0 Z}(M_A^2)
\label{DabelsteinCond}
\end{align}
on the renormalization constant $\delta \tan\beta\equiv\dTB$. Here
``pure divergence'' denotes a term of the order
$\Delta=\frac{2}{4-D}-\gamma_E+\log4\pi$ in dimensional reduction and
$\Sigma_{A^0 Z}$ is the unrenormalized $A^0Z$ two-point
function. These schemes have the advantage of defining $\tan\beta$ 
in a universal way in the Higgs sector and being
technically very convenient. On the other hand, they do not
imply any obvious relation between $\tan\beta$ and observable
quantities; hence, they might lead to a gauge dependence of
$\tan\beta$. 

The gauge dependence of $\tan\beta$ can be computed by using an extended
Slavnov-Taylor identity introduced in \cite{Kluberg,PiSiSTI}:
\begin{align}
\tilde{S}(\Gamma)\equiv S(\Gamma)+ \chi\, \partial_\xi\Gamma & = 0.
\label{eq:ExtSTI}
\end{align}
Here $\xi$ denotes an arbitrary gauge parameter in the
gauge-fixing term and $\chi$ is a fermionic variable acting as the BRS
transformation of $\xi$, and $S(\Gamma)$ is the usual Slavnov-Taylor
operator. The validity of $\tilde{S}(\Gamma)=0$ is equivalent to the
gauge independence of all input parameters, in particular of
$\tan\beta$.

Consider the $\overline{DR}$-scheme as an example. In this scheme, the
renormalization constant $\dTB$ is obviously gauge independent:
\begin{align}
\partial_\xi\dTB^{\rm fin} & = 0.
\label{DRBarContra2}
\end{align}
But the validity of $\tilde{S}(\Gamma)=0$ would imply a certain
gauge-parameter dependence at the one-loop level:
\begin{align}
\partial_\xi\dTB^{\rm fin} & \propto
(-\sin\beta A_1 + \cos\beta A_2),
\label{DRBarContra1}
\end{align} 
where $A_i=\Gamma^{\rm (1),reg}_{\chi Y_{\phi_i}}$ denotes the
unrenormalized one-loop Green function with the BRS transform $\chi$
of the gauge parameter and the source $Y_{\phi_i}$ of the BRS
transformation of the Higgs field $\phi_i$. The Green functions $A_i$
can be calculated, but their results depend on the choice of the gauge
fixing. In the class of $R_\xi$-gauges, 
\begin{align}
(A_1, A_2)\propto(\cos\beta,\sin\beta),
\end{align}
and thus the r.h.s.\ of (\ref{DRBarContra1}) yields zero and 
is compatible with (\ref{DRBarContra2}). This shows that $\tan\beta$
is gauge independent in the $\overline{DR}$-scheme at the one-loop
level and in the class of $R_\xi$-gauges. However, in a
non-$R_\xi$-gauge, where the physical $A^0$ boson is introduced into
the gauge-fixing function
\begin{align}
\label{GeneralizedGauge}
\F^Z & =  \partial_\mu Z^\mu + M_Z(\xi G^0 + \zeta^{ZA^0}A^0),
\end{align}
virtual $A^0$ bosons contribute to $A_{1,2}$ instead of virtual
Goldstone bosons, and the results for $A_{1,2}$ are modified:
\begin{align}
(A_1, A_2)\propto(-\sin\beta,\cos\beta).
\end{align}
Therefore, in this gauge the r.h.s.\ of (\ref{DRBarContra1}) is
non-vanishing and thus in contradiction with the
${\overline{DR}}$-condition (\ref{DRBarContra2}). In other words, in
general gauges, the $\overline{DR}$-scheme leads to a violation of
$\tilde{S}(\Gamma)=0$ and to a gauge dependence of $\tan\beta$ already
at the one-loop level. In addition, in \cite{Yamada01} it was found
that the $\overline{DR}$-scheme leads to a gauge dependence at the
two-loop level even in the $R_\xi$-gauges.

The gauge dependence of $\tan\beta$ in the DCPR-schemes can be studied
in a similar way. It turns out that these schemes lead to a gauge
dependence already in $R_\xi$-gauges at the one-loop level.

\section{Gauge- and process-independent schemes and their drawbacks}

As an alternative to the gauge-dependent schemes discussed in the
previous section, we consider now a class of gauge-independent
schemes. Three examples are
\begin{align}
&\mbox{Tadpole scheme:}& \delta t_\beta^{\rm fin}&\stackrel{!}{=}
 \mbox{const.}\left(\frac{\delta t_1}{v_1} + \frac{\delta t_2}{v_2}
 \right),
\label{Tadpole}
\\
&\mbox{$m_3$-scheme:}&
 \delta m_3^{\rm fin}&\stackrel{!}{=} 0,
\label{m3Scheme}
\\
&\mbox{HiggsMass-scheme:}&
 \cos^2(2\beta)&\stackrel{!}{=}\frac{M_h^2 M_H^2}{M_A^2(M_h^2 +
 M_H^2 - M_A^2)}.
\label{HiggsMass}
\end{align}
Each of these schemes has a different underlying intuition. In the
first case, $\tan\beta$ is defined in a minimal gauge-independent way
via a relation to tadpole counterterms,
in the second case an indirect definition via the soft-breaking
parameter $m_3$ is used, and in the third case $\cos(2\beta)$ and
$\tan\beta$ are defined in terms of a ratio of physical Higgs masses.

It can be shown that these three schemes are in agreement with
$\tilde{S}(\Gamma)=0$ and thus with the gauge independence of
$\tan\beta$. Furthermore, they define $\tan\beta$ in a universal way,
using only quantities of the MSSM Higgs sector.

Unfortunately, in spite of these advantageous properties all three
schemes are not useful in practice because they cause very large
numerical uncertainties in loop corrections to quantities involving
$\dTB$. This can be exemplified by the renormalization-scale
dependence of $\tan\beta$ in the first two schemes, (\ref{Tadpole}),
(\ref{m3Scheme}). 
\begin{table}[bt]
\begin{tabular}{cc}
\begin{tabular}{|l|r|r|r|}
\hline
              & $\overline{DR}$ &(\ref{Tadpole})   &(\ref{m3Scheme})    \\
$\tan\beta=3$ & $-0.1$ & {4.5}       & {0.8}   \\
$\tan\beta=50$& $-0.2$ & {370.7}     & {285.3}\\
\hline
\end{tabular}
&
\begin{tabular}{|c|c|c|c|c|}
\hline
 $\overline{DR}$ & DCPR & (\ref{Tadpole}) & (\ref{m3Scheme}) & (\ref{HiggsMass}) \\
134.6  & 134.4 & {173.5} & {143.2} & {119.6} \\
\hline
\end{tabular}\\
(a) & (b)
\end{tabular}
\caption{(a): The renormalization-scale $\bar\mu$-dependence
$\partial\tan\beta/\partial\log\bar\mu$ of the schemes (\ref{Tadpole}),
(\ref{m3Scheme}) in comparison with the $\overline{DR}$-scheme.  We have chosen
$M_A=500$ GeV, and the remaining parameter values are chosen according
to the $M_h^{\rm max}$-scenario of \cite{Scenarios}.
(b): Results for the one-loop corrected lightest Higgs mass $M_h$ using the
same parameters and $\tan\beta=3,{\bar\mu}=m_t$.}
\label{TableMuDep}
\end{table}
Table \ref{TableMuDep}(a) shows that while the renormalization-scale
dependence in the $\overline{DR}$-scheme is quite modest, the one in
the schemes (\ref{Tadpole}), (\ref{m3Scheme}) can be extremely large,
and hence leads to an inacceptable numerical instability of loop
calculations. Although $\tan\beta$ in the HiggsMass-scheme is
renormalization-scale independent, it involves numerically very large
contributions to observable quantities and also leads to numerical
instabilities. As an example, table \ref{TableMuDep}(b) shows the results
for the one-loop corrected mass of the lightest MSSM Higgs boson,
$m_h$. Obviously, the results obtained in the schemes
(\ref{Tadpole}--\ref{HiggsMass}) deviate strongly from the results
obtained in the $\overline{DR}$- or DCPR-schemes.

As shown in \cite{TB}, one can generalize from these three particular
schemes to the class of all schemes where $\tan\beta$ is defined via
quantities of the MSSM Higgs sector (i.e.\ $\dTB$ is composed of Higgs
self energies and tadpoles). In all gauge-independent schemes of this
class numerical instabilities like the ones shown in table
\ref{TableMuDep} appear, so unfortunately all these schemes are
useless in practice.

\section{Process-dependent schemes}

The results obtained up to now are negative. In the considered classes
of schemes, there is no scheme that combines all three desirable
properties
\begin{itemize}
\item gauge independence,
\item numerical stability,
\item process independence.
\end{itemize}
The $\overline{DR}$- and DCPR-schemes are gauge dependent,
and the gauge- and process-independent schemes discussed in the
previous section are numerically unstable. As an alternative we can
think about dropping the requirement of process independence as
advocated in \cite{CGGJS96}. Two possible schemes are given by
\begin{align}
\tan^2\beta&\stackrel{!}{=}
\mbox{const.}\times\Gamma(A^0\to\tau\tau),\\
\tan^2\beta&\stackrel{!}{=}
\mbox{const.}\times\Gamma(H^+\to\tau^+\nu),
\end{align}
where ``const.'' denotes the kinematical prefactors of the decay
widths. Since $\tan\beta$ is directly related to observables in this
way it is gauge independent, and these schemes do not induce numerical
instabilities. However, these schemes have other disadvantages.
At first they are technically complicated since the evaluation of
$\dTB$ requires the calculation of the full decay widths. In
particular, the second process 
involves infrared divergent QED-corrections that cannot be split off
from the definition of $\tan\beta$. Furthermore, such
process-dependent schemes introduce a flavour dependence, which seems
unnatural since these decays are just two examples amongst a variety
of potential observables for the experimental determination of
$\tan\beta$. 

\section{Conclusions}

The gauge- and process-independent schemes presented are practically
useless because of their numerical instabilities. The well-known
$\overline{DR}$- and DCPR-schemes are gauge dependent already at the
one-loop level, but they can be used very well in practice. Among these
schemes the $\overline{DR}$-scheme is preferable for two
reasons. It is gauge independent at the one-loop level in the class of
$R_\xi$-gauges, and a recent study has shown that its numerical
behaviour is particularly stable \cite{FHHW}. An alternative is
provided by process-dependent schemes. Among these, the decay
$A^0\to\tau\tau$ is advantageous since there the infrared divergent
QED-corrections can be split off. However, as all process-dependent
schemes, this scheme is technically relatively complicated and defines
$\tan\beta$ in a non-universal way. Therefore we assess the
$\overline{DR}$-scheme as the best renormalization scheme for
$\tan\beta$. It is technically the easiest, numerically the most
well-behaved, and it is still gauge independent in the practically most
important case.

\begin{flushleft}

\end{flushleft}

\end{document}